# Classification and Powerlaws: The Logarithmic Transformation

Loet Leydesdorff [1] & Stephen Bensman [2]

**Abstract**

Logarithmic transformation of the data has been recommended by the literature in the case of highly skewed distributions such as those commonly found in information science. The purpose of the transformation is to make the data conform to the lognormal law of error for inferential purposes. How does this transformation affect the analysis? We factor analyze and visualize the citation environment of the *Journal of the American Chemical Society* (*JACS*) before and after a logarithmic transformation. The transformation strongly reduces the variance necessary for classificatory purposes and therefore is counterproductive to the purposes of the descriptive statistics. We recommend against the logarithmic transformation when sets cannot be defined unambiguously. The intellectual organization of the sciences is reflected in the curvilinear parts of the citation distributions, while negative powerlaws fit excellently to the tails of the distributions.

[1] Amsterdam School of Communications Research (ASCoR), University of Amsterdam, Kloveniersburgwal 48, 1012 CX Amsterdam, The Netherlands; loet@leydesdorff.net, http://www.leydesdorff.net.
[2] LSU Libraries, Louisiana State University, Baton Rouge, LA 70803-3300, USA; notsjb@lsu.edu.

# 1. Introduction

The problem under analysis in this paper has its genesis in a controversy that erupted on the pages of *JASIST* over the use of the Pearson correlation coefficient as a similarity measure in author cocitation analysis (ACA). Ahlgren *et al.* (2003) challenged basing ACA on the Pearson *r* with the argument that this measure is sensitive to zeros in the sense that the relationships among the authors change when authors not citing any of them are added to the set. These authors proposed alternative measures such as the cosine. White (2003) defended the method of the Drexel school (White & Griffith, 1981, 1982; McKain, 1990) by showing that the Pearson *r* and the cosine lead to similar classification and mapping results when using Ahlgren *et al.*'s own data.[3]

The Pearson *r* is a measure of the closeness of the fit of observation points to a regression line and is therefore a linear statistical model. Linear statistical models rely upon a number of basic assumptions. Without these assumptions, the data for them must be mathematically transformed so that the condition of linearity is satisfied. In a review of the key literature on such transformations, Hoyle (1973) summarized the assumptions conditional to the use of linear models as follows:

   (a) additivity—that is, the main effects combine linearly to "explain" the

---

[3] The author cocitation matrix is a co-occurrence matrix (Van Rijsbergen, 1977). This symmetrical matrix can directly be used as a proximity matrix, for example, for the purpose of multi-dimensional scaling (Leydesdorff & Vaughan, forthcoming: Vaughan & You, 2005). The co-occurrence matrix is based on the multiplication of the original matrix of documents with citations (or other textual elements) as its attributes with the transposed of this matrix (Leydesdorff, 1989; Engelsman & Van Raan, 1991). Comparison of the observed values in a co-occurrence matrix with the expected ones can be elaborated into a statistics by using the chi-square method (Michelet, 1988; Zitt *et al.*, 2000).

observations;

(b) constant variance—that is, the observations are assumed to have a constant variance about their varying means. Explicitly this means that the variance is independent of both the expected value of the observations and the sample size;

(c) normality—that is, the observations are assumed to have a normal distribution. (p. 203)

For their part, Box and Cox (1964, 211) further qualified the assumptions underlying linear statistical models by adding to them simplicity of model structure and independence of observations.

Information science data rarely allow for the satisfaction of these assumptions. This is particularly true of scientific journal citation data, due to the structure of scientific journal sets even after an initial classification process, and the stochastic processes underlying the distributions resulting from this structure. If the data is heavily skewed—like it is often the case in information science—one should consider to perform a logarithmic transformation. Logarithmically transformed data may exhibit log-normality, and thus allow for using the Pearson correlation coefficient.

In this study, we lognormalize journal-journal citation data before using the Pearson correlation (as an initial step in factor analysis). Might this transformation provide an option for testing different possible classifications of journals for their significance (Leydesdorff, forthcoming)? We found that the logarithmic transformation did not add clarity to the classificatory process. This accords with White's (2004, p. 844) expectation that "if *mapping* the correlation data is the goal, one merely wants the *r*'s to reflect degrees of similarity among the authors, and so significance tests from inferential statistics are not (I would think) of primary interest."

We shall show below that the logarithmic transformation even worsens the quality of the classification. These results raise the question of the role of inferential statistics and the logarithmic transformation in the mathematical and statistical classification of observations into sets. We explore this question in this study by combining the theoretical background with empirical tests. In short, we will explain why the logarithmic transformation is counterproductive to the objective of classification in the case of bibliometric data (which typically exhibit heavily skewed distributions). This conclusion has implications for the interpretation and use of powerlaws in bibliometric data (Katz, 2000).

**2. Statistics and information science**

In terms of their underlying subject structure scientific journal sets are governed by two bibliometric laws: Bradford's Law of Scattering and Garfield's Law of Concentration. The first was posited by Bradford (1934, at p. 86), the director of the Science Museum Library in London, as a result of bibliographic studies done at this library. The second law was formulated by Garfield (1971) in the context of the selection of journals for inclusion in the *Science Citation Index (SCI).* The implications of these insights for information science were elaborated by Brookes (1977, 1979, 1980a, 1980b, 1984; Brookes & Griffith, 1979).

*2.1    Scattering and concentration of journal sets*

Bradford (1934) analyzed the distribution of articles in two subject areas: Applied Geophysics, 1928-1931, and Lubrication, 1931-1933. In neither area was he able to determine the number of journals that had no articles on the topics but potentially could, stating:

> …the number of journals which contain journals on the subjects in question is of the order of a thousand. But the periodicals themselves could not be specified without scrutinizing a much larger number of periodicals during a long period. And even when the actual producers during a period of years had been ascertained, new sources would certainly appear during a further period. It follows that the only way to glean all the articles on these subjects would be to scrutinize continually thousands of journals, the bulk of which would only yield occasional references or none at all. (p. 86)

In other words, Bradford's Law states that the distribution of articles on a given scientific topic over a set of journals is such that a large proportion of these articles appear in a relatively small core set of journals, while the remaining articles are spread over zones of journals that must increase exponentially in numbers of titles to obtain the same number of articles on the topic as in the core. Due to Bradford's Law, unambiguously delineated ("crisp") subject sets of scientific journals cannot be expected, and the purpose of the initial classification process is merely to approximate such subject sets as closely as practicable (Bensman, 2000; 2001; Zadeh, 1965).

The composite and multidisciplinary nature of science underlies also Garfield's Law of Concentration, which Garfield (1971) considered as the citation corollary of Bradford's Law of Scattering. Garfield (1971; 1972; 1983, 21-23 and 158-163) developed his law as a result of an analysis of references published during the last quarter of 1969 in the 2,200 journals then covered by the *SCI*. He found a distribution similar to the one discovered by Bradford because citations in an individual discipline like chemistry concentrate on a small core of journals. The ubiquity of such disciplinary cores caused Garfield to reformulate Bradford's Law by transposing it from the level of individual disciplines to the level of science as a whole. Likening Bradford's Law to a comet with the core journals of a discipline representing the nucleus and the zones acting as the tail, Garfield posited that the tail of the literature of any given scientific discipline consists in large part of the nuclei or cores of the literatures of other disciplines. Thus, a multidimensional space is spanned in terms of a variety of core sets, but each core includes a large part of the others in the tail of the distribution. According to Garfield, this phenomenon causes citations to concentrate on a small multidisciplinary core of some 500 to 1,000 journals representing all of science.

On the basis of these two laws, one cannot expect that scientific journal sets will be homogeneous in terms of their subject matter. A journal set defined by a given scientific discipline can be comprised of subsets of journals which can be classed in the sub-disciplines of this discipline as well as subsets of journals from other disciplines that contain materials of interest to the defining discipline. This latter subset can be considered a partial subset because it also contains materials not pertinent to the defining discipline. Moreover, a scientific journal set can also be broken down into subsets by criteria other than subject ones such as nationality, language, type of publisher, or purpose, e.g., research, review, informational, and instructional.

The composite structure of scientific journal sets dictates that their data distributions are for the most part compound ones. A compound distribution can be defined as a type of probability distribution arising when a parameter of the distribution such as the arithmetic mean is itself a random variable with its own probability distribution (Everitt, 1998, 71). Scientific journal distributions result from the Poisson process, which is the random occurrence of events such as citations over continuums of time and space. For these distributions space is defined in terms of the subsets comprising the set. Each of the subsets of a scientific journal set has different underlying probabilities and therefore a different expected value or arithmetic mean.

Two stochastic processes govern these scientific journal distributions. The first is heterogeneity. The variances around the arithmetic means tend to vary in proportion to the size of the arithmetic means, thereby violating one of the basic assumptions of linear statistical models. The second stochastic process is contagion. A term first suggested by the study of the probability distributions of epidemics, contagion became more broadly used to designate situations where trials are not independent, because the occurrence of an event affects the probability of its further occurrence. Citations act in such a manner, since each citing of a journal increases its probability of being cited again. This has been discussed in science studies as the Matthew effect (Merton, 1968) and more recently as the mechanism of preferential attachment which is well-known for generating negative powerlaws (Barabási, 2002; Barabási *et al*., 2002; Katz, 1999, 2000; Wagner & Leydesdorff, forthcoming). The linear fit of a log-log distributional chart can be used as a test for this preferential attachment mechanism.

Both heterogeneity and contagion act multiplicatively instead of additively, creating exponential and curvilinear relationships instead of the assumed additive, linear ones. Feller (1943) proved that heterogeneity and contagion serve as the basis for two different models of the negative binomial distritbution (NBD). Therefore, the NBD could serve as a probabilistic model of the causal processes in scientific journal distributions. The NBD can be normalized by the arc-sinh transformation (Anscombe, 1948). However, these precise mathematical probability models require crisp sets, which cannot be expected to exist in scientific journal data given Bradford's and Garfield's Laws.

*2.2    The logarithmic transformation*

As a result of their structure, scientific journal subject sets contain data unrelated to the subject, causing extreme statistical outliers that distort parameter estimates and prevent precise mathematical fits to theoretical curves. However, these outliers are meaningful because they span the structure in the data. They are indicated by the variance, but much less so by the arithmetic mean. Consequently, the latter is not an accurate measure under these circumstances. The vast majority of science journal distributions have a variance significantly much greater than their arithmetic mean.

In a landmark article Bartlett (1947, 43) specified the significance of this phenomenon in terms of the dynamics of a biological system. According to him, the natural explanation of a variance greater than the mean is that the mean level itself fluctuates. He noted that for biological populations, increases in numbers are often proportional to the numbers already present, giving rise to variations in the mean from place to place themselves proportional to the local mean. In the case of a variance greater than the mean, the literature advises considering a logarithmic transformation of the data (Bartlett, 1947; Quenouille, 1950). For his part, Elliott (1977, 33) considered the variance being greater than the mean as a sign of the negative binomial distribution, and he made the following recommendations: 1) with no zero counts, simple logarithmic transformation of the data; 2) with some zero counts, add one to the observations before performing the logarithmic transformation. Quenouille (1950, 165) stated that the logarithmic transformation tends to restore normality in the distribution and equalize the variances simultaneously, whereas Hoyle (1973, 207) cites a number of studies empirically showing the logarithmic transformation as a way of making the data conform to the three linear-model assumptions of additivity, constant variance, and normality.[4]

In summary, the logarithmic transformation of data enables the analyst to switch the law of error for tests of significance in linear models from the normal distribution to the lognormal distribution. In their book on the latter distribution Aitchison and Brown (1957) defined the lognormal distribution as "the distribution of a variate whose logarithm obeys the normal law of probability" (p. 1). According to them, many of the properties of the lognormal may be immediately derived from those of the normal distribution.

---

[4] Bensman (1996) and Bensman and Wilder (1998) found that the logarithmic transformation induced not only normality in the data but also that the semi-logarithmic model of multiple regression, where only the dependent variable is logarithmically transformed, eliminated severe heteroscedasticity.

Aitchison and Brown believed that the lognormal distribution was as fundamental a distribution in statistics as the normal distribution: "It arises from a theory of elementary errors combined by a multiplicative process, just as the normal distribution arises from a theory of elementary errors combined by addition" (pp. 1-2). Keynes (1921, 198-200) regarded as the main advantage of the lognormal distribution the possibility it offered of adapting without much trouble to asymmetrical phenomena numerous expressions which had already been calculated for the normal law of error. In contrast to the normal distribution, which is centered on the arithmetic mean, the lognormal distribution is centered on the geometric mean, which can be calculated by first calculating the arithmetic mean of the logarithmically transformed data and then taking this mean's antilogarithm. Thus, we can see that the purpose of the logarithmic transformation is to create a model that conforms to the requirements of the normal law of error for inferential purposes. It does this by artificially reducing the amount of variance to that of the normal distribution.

*2.3    The implications for information science and technology*

In a series of papers B. C. Brookes worked out the deeper implications of the logarithmic transformation for information science. In the first of this series, Brookes (1977) came to the conclusion that Bradford had succeeded in formulating an empirical regularity, which has pure and hybrid forms, but that all the variants can be subsumed under a simple logarithmic law which escapes exact expression in conventional frequency terms. In this analysis he closely linked Bradford's Law with set definition, insisting upon the need for homogeneity of the data. Brookes (pp. 194-197) stated that most Bradford anomalies are due to inhomogeneous data, and he characterized *SCI* citation data specifically as inhomogeneous.

Utilizing the logarithmic Law of Anomalous Numbers advanced by Benford (1938), Brookes developed Bradford's Law into a linear model of social reality with the type of deviations from linearity indicating the nature of the stochastic process that is occurring. On the basis of this model he developed a new theory of frequency-rank statistics especially applicable to social analysis. Brookes and Griffiths (1979) noted that in many social contexts, when a homogeneous ensemble of sources has been engaged in some discrete homogeneous activity, ranking the sources in descending order by frequency counts results in a distribution that is logarithmic. Brookes (1979) came thus to regard Bradford's Law as a new calculus for the social sciences.

Brookes (1980a, 219-220; 1980b) found the negative binomial to be the standard statistical distribution that fits Bradford data, and he argued that information quantities *should* hence be measured logarithmically. The logarithmic transformation was thus made central to the description of the data. Generalizing his theory, Brookes (1984) proved that Bradford's Law was almost identical to other empirical bibliometric laws such as those of Lotka, Zipf, and Price, and he formulated an equation which he called "the empirical Log Law" for calculating rank distributions. However, the issue of whether one should obey this "empirical law" logarithmically transforming citation data before analysis has remained unresolved in empirical research (e.g., Drott & Griffith, 1978; Egghe & Rousseau, 2003). Let us put this recommendation to the test.

**3. Methods and materials**

*3.1    Data*

The role of inferential statistics and logarithmic transformation in numerical classification and mapping will be analyzed in terms of the allocation of scientific journals into different subject sets. Our data was collected from the CD-Rom version of the *Journal Citation Reports 2003* of the *Science Citation Index*. We included all journals which provide more than one percent of the citations to articles in the *Journal of the American Chemical Society* during this year (Leydesdorff & Cozzens, 1993). This leads to the demarcation of the set of 21 journals listed in Table 1.

**Table 1.** Library of Congress Subject Headings and Class Groups for the 21 Journals Citing the *Journal of the American Chemical Society*

| Titles | Publishers | Subject Headings | Call Number | Class Group | Class Group Hierarchy |
|---|---|---|---|---|---|
| *Science* | American Association for the Advancement of Science | 1. Science. | Q1 | Science (General) | Science (General) |
| *Angewandte Chemie-International Edition* | Wiley-VCH (1) | 1. Chemistry. | QD1 | Chemistry | Chemistry |
| *Chemical Communications (2)* | Royal Society of Chemistry | 1. Chemistry. | QD1 | Chemistry | Chemistry |
| *Chemistry-A European Journal* | VCH Verlagsgesellschaft (1) | 1. Chemistry. | QD1 | Chemistry | Chemistry |
| *Chemical Reviews* | American Chemical Society | 1. Chemistry. | QD1 | Chemistry | Chemistry |
| *Journal of the American Chemical Society* | American Chemical Society | 1. Chemistry. | QD1 | Chemistry | Chemistry |
| *Dalton Transactions (3)* | Royal Society of Chemistry | 1. Chemistry, Inorganic. 2. Chemistry, Physical and theoretical. | QD146 | Inorganic chemistry | Chemistry--Inorganic chemistry |
| *Inorganic Chemistry* | American Chemical Society. | 1. Chemistry, Inorganic. 2. Bioinorganic chemistry . | QD146 | Inorganic chemistry | Chemistry--Inorganic chemistry |
| *Journal of Organic Chemistry* | American Chemical Society | 1. Chemistry, Organic. | QD241 | Organic chemistry | Chemistry--Organic chemistry |
| *Organic and Biomolecular Chemistry (4)* | Royal Society of Chemistry | 1. Chemistry, Organic. 2. Bioorganic chemistry. 3. Chemistry, Physical organic | QD241 | Organic chemistry | Chemistry--Organic chemistry |
| *Tetrahedron* | Pergamon Press | 1. Chemistry, Organic. | QD241 | Organic chemistry | Chemistry--Organic chemistry |
| *Tetrahedron Letters* | Pergamon Press | 1. Chemistry, Organic. | QD241 | Organic chemistry | Chemistry--Organic chemistry |

| *Organic Letters* | American Chemical Society | 1. Chemistry, Organic. | QD241 | Organic chemistry | Chemistry—Organic chemistry |
| --- | --- | --- | --- | --- | --- |
| *Macromolecules* | American Chemical Society | 1. Macromolecules. 2. Polymers. 3. Polymerization. | QD380 | Polymers. Macromolecules | Chemistry--Organic chemistry--Polymers. Macromolecules |
| *Journal of Organometallic Chemistry* | Elsevier Sequoia | 1. Organometallic compounds . | QD410 | Organometallic chemistry and compounds | Chemistry--Organic chemistry--Organometallic chemistry and compounds |
| *Organometallics* | American Chemical Society | 1. Organometallic compounds. | QD410 | Organometallic chemistry and compounds | Chemistry--Organic chemistry--Organometallic chemistry and compounds |
| *Journal of Chemical Physics* | American Institute of Physics | 1. Chemistry . 2. Physics 3. Chemistry, Physical and theoretical. | QD450 | Physical and theoretical chemistry | Chemistry--Physical and theoretical chemistry |
| *Journal of Physical Chemistry A (5)* | American Chemical Society | 1. Chemistry, Physical and theoretical . | QD450 | Physical and theoretical chemistry | Chemistry--Physical and theoretical chemistry |
| *Journal of Physical Chemistry B (5)* | American Chemical Society | 1. Chemistry, Physical and theoretical . | QD450 | Physical and theoretical chemistry | Chemistry--Physical and theoretical chemistry |
| *Langmuir* | American Chemical Society | 1. Surface chemistry. 2. Colloids. 3. Surfaces (Physics). | QD506. | Surface chemistry | Chemistry--Physical and theoretical chemistry--Surface chemistry |
| *Biochemistry-US* | American Chemical Society | 1. Biochemistry. | QP501 | Animal biochemistry | Physiology--Animal biochemistry |

(1) A journal of the Gesellschaft Deutscher Chemiker.
(2) Title changed in 1996 from: *Journal of the Chemical Society. Chemical Communications.*
(3) Formed by the union in 2000 of *Journal of the Chemical Society*, *Dalton Transactions,* and *Acta Chemica Scandinavica* to become *Dalton,* which in 2003 became *Dalton Transactions*.
(4) Formed in 2003 by the union of *Perkin 1* and *Perkin 2*. *Perkin* was formed in 2000 by the merger of: *Journal of the Chemical Society. Perkin Transactions 1*;, and part of *Acta Chemica Scandinavica*. *Perkins 2* was formed in 2000 by the merger of *Journal of the Chemical Society. Perkin Transactions II*, and part of *Acta Chemica Scandinavica*.
(5) Continues in part as *Journal of Physical Chemistry* since 1997.

One interesting feature of these journals is their publisher structure. Most of these journals are either published by scientific societies or associated with scientific societies. Thus, eleven are published by the American Chemical Society; three are published by the Royal Society of Chemistry; one by the American Association for the Advancement of Science; one by the American Institute of Physics; and two are journals of the Gesellschaft Deutscher Chemiker even though issued by commercial publishers. Society journals are the ones most highly rated by chemists and used in chemistry libraries (Bensman, 1996). Citations concentrate on both journals of scientific societies and elite research programs, showing that scientists from these programs publish in society journals (Bensman & Wilder, 1998). Thus, the publisher structure of the 21 journals is evidence that these journals rank high in the social structure of chemistry and are a manifestation of the intercommunication pattern of the chemistry scientific elite.

The set structure of the database will first be analyzed by the logical method of induction and analogy set forth by Keynes (1921). This can be done by showing what subject headings and class numbers are assigned to these 21 journals by the United States Library of Congress (LC). Table 1 gives these subject headings and class numbers. The subject headings should be self-evident, but the class numbers may require some explanation. In the standard work on LC Classification, Chan (1999, p. 12-16) states that the LC scheme is based on "literary warrant." A classification scheme based on literary warrant is not logically deduced from some abstract philosophical system for classifying knowledge but inductively developed in reference to the holdings of a particular library or to what is or has been published. In other words, it is based on what the actual literature of the time warrants. Each of the individual schedules was initially drafted by LC subject specialists, who consulted bibliographies, treatises, comprehensive histories, and existing classification schemes to determine the scope and content of an individual class and its subclasses. The LC has a policy of continuous revision to take current literary warrant into account, so that new areas are developed and obsolete elements are removed or revised.

Analysis of the class numbers shows that the 21 journals have been classified logically into three basic subclasses or sets. Thus, the journal *Science* is classed in Q1 or Science (General). It is followed by 19 journals that are classed within QD or Chemistry and its hierarchical subclasses. The last journal, *Biochemistry-US,* has been classed within the subclass Animal Biochemistry within the subclass QP or Physiology. Thus, the conclusion from the logical LC classification of this citation environment of the *JACS* is that we dealing with a core of 19 journals fully within the chemistry set and two journals—*Science* and *Biochemistry-US*—only partially within the chemistry set. However, given Bradford's and Garfield's Laws, even the 19 journals of the chemistry core can be expected to be only partially within the chemistry set and as to have facets outside this set.

*3.2     Methods*

A matrix of 21 x 21 cells can be constructed from the list of journals provided in Table 1 (Appendix I). This matrix is asymmetrical: the cases (rows) are cited by the same set of journals in the columns. The descriptive analysis of the subject relationships among the 21 journals of the database will first be done in terms of the frequency with which each of the journals was cited in 2003 by the journals of the database. After two sections with descriptive statistics, we shall proceed to the (Q-)factor analysis of the aggregated citation matrix among the 21 journals in order to find communalities in their being-cited patterns. Varimax rotation and Kaiser maximalization on the basis of the Pearson correlation matrix will be used. The results are visualized using the Pearson correlation matrix as input to the algorithm of Kamada & Kawai (1989)[5] as available in Pajek.[6] The data matrix is thereafter transformed by taking the logarithm of the values in the cells, and the analysis is then repeated. Because the citation matrix contains some zeros and $\log(0) = -\infty$, 1 was added to all values in this pass (Elliott, 1977, 33).

The vector-space model based on the cosine (Salton & McGill, 1983) is more suitable for the visualization since the cosine runs from 0 to 1, while the Pearson correlations can vary from $-1$ to $+1$.[7] The two similarity measures are otherwise equivalent (Jones & Furna, 1987). Since the matrix under study did not contain many zeros (cf. Ahlgren *et al*., 2003), and given our research focus on the effects of the logarithmic transformation on the normality and/or lognormality of the distribution, we shall use the Pearson correlation exclusively as the basis of both the statistics and the visualizations.

---

[5] This algorithm represents the network (that is, the matrix) as a system of springs with relaxed lengths proportional to the edge length. Nodes are iteratively repositioned to minimize the overall 'energy' of the spring system using a steepest descent procedure. The procedure is analogous to some forms of non-metric multi-dimensional scaling.

[6] Pajek is freely available for non-commercial purposes at http://vlado.fmf.uni-lj.si/pub/networks/pajek/ .

[7] If one includes the negative values of Pearson correlations, these can be visualized using Pajek as dashed lines, but then it is no longer possible to show the structure in the correlation in a single picture. Therefore, we will use the Pearson correlations in the visualizations only insofar as the values of *r* are larger than or equal to zero. This procedure usually provides sufficient information for illustrating the factor structure with a corresponding visualization (Leydesdorff, 1987; Leydesdorff & Cozzens, 1993).

## 4. Results

*4.1     The effects of the logarithmic transformation on the distributions*

To begin the analysis, the shape of the frequency distributions of the citing journals and the effect of the logarithmic transformation on this shape will be shown in detail for two of the journals, the *Journal of the American Chemical Society* (*JACS*) and *Science*.  The first is the linchpin of the database's chemistry set; the second has been logically classified above as being outside this chemistry set.  Figures 1 and 2 graph the shapes of the distributions for these journals in both the raw-count and logarithmic form. These histograms were constructed by dividing the range of the citations into deciles and then grouping the citing journals by these deciles.

Figure 1. *Journal of the American Chemical Society* Distributions

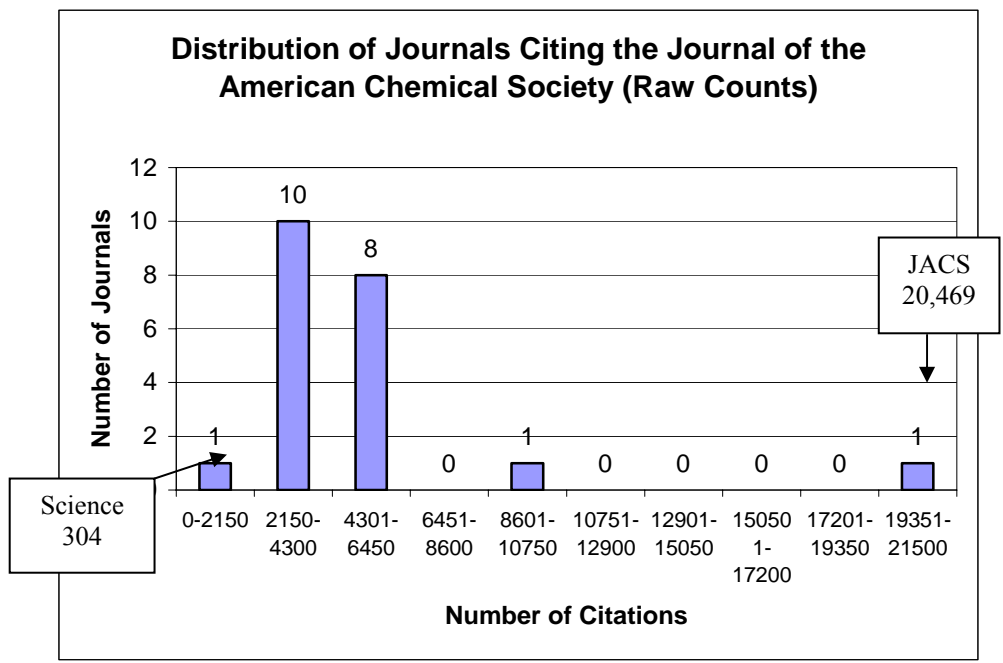

Arithmetic Mean = 4,744.95
Variance = 162,772,608.55
Variance-to-Mean Ratio = 3,429.46
Type of Distribution: Compound Poisson, Contagious

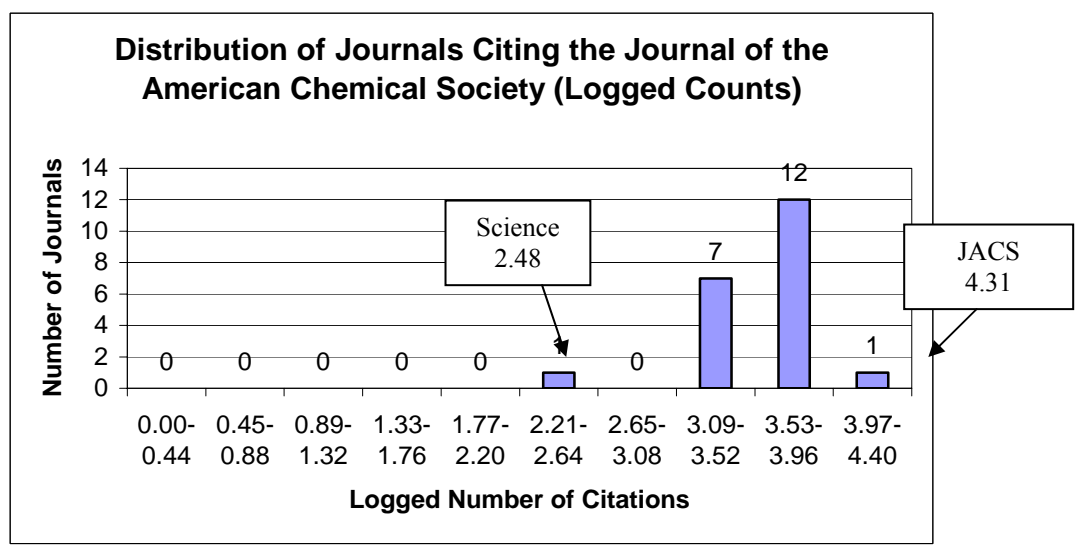

Arithmetic Mean = 3.57
Variance = 0.11
Variance-to-Mean Ratio = 0.03
Type of Distribution: Lognormal

Figure 2. *Science* Distributions

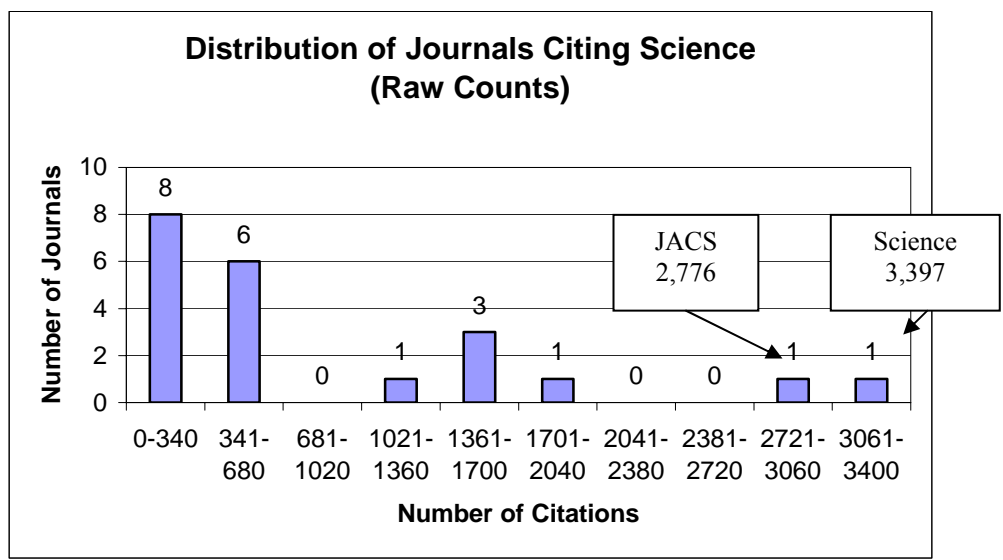

Arithmetic Mean = 863.57
Variance = 836,520.76
Variance-to-Mean Ratio = 968.68
Type of Distribution: Compound Poisson, Contagious

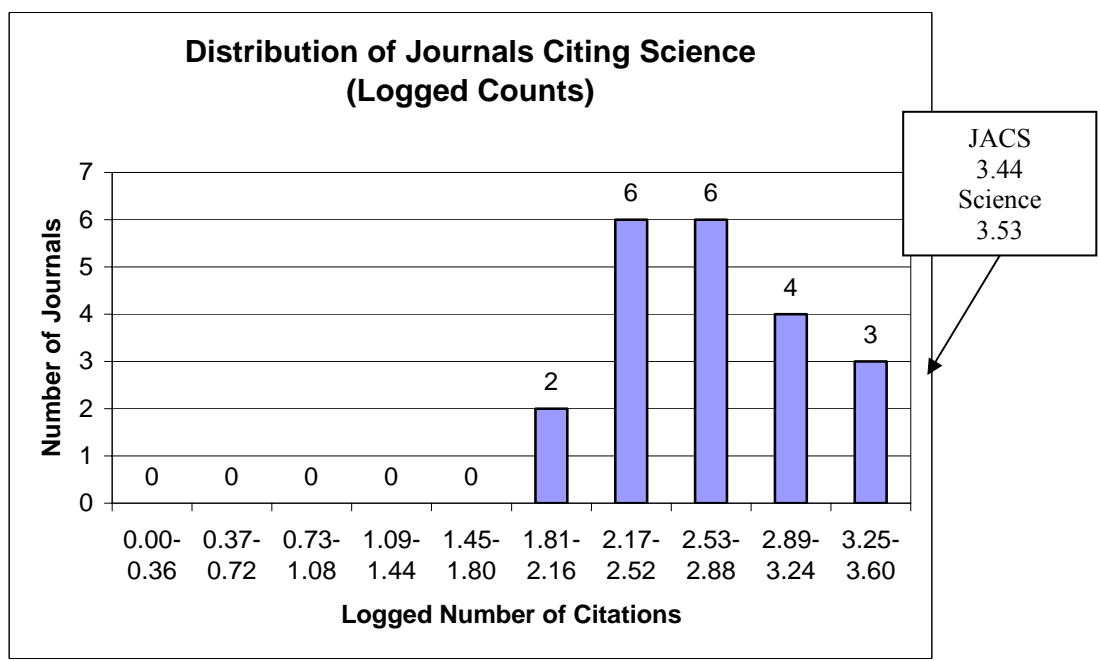

Arithmetic Mean = 2.72
Variance = 0.21
Variance-to-Mean Ratio = 0.08
Type of Distribution: Lognormal

In both cases it is clear that the top journals citing these two journals were themselves—with *JACS* having 20,469 self-citations and *Science* having 3,397 self-citations. It can be deduced that the bulk of the *Science* self-citations were not to chemistry articles. This can be seen in the imbalance with which these two journals cited each other. Thus, *Science* was the lowest of the journals citing *JACS,* with a count of only 304, whereas *JACS* was the second-highest of the journals citing *Science,* with a count of 2,776. In the raw-count form both journals' distribution manifest the typical shape of a compound Poisson, contagious distribution with the majority of the journals concentrated below the arithmetic mean, the long tail to the right causing huge variance, and an extremely high variance-to-mean ratio—3,429.46 for *JACS* and 968.68 for *Science*. These shapes and high variance-to-mean ratios are natural products of the probabilistic heterogeneity of the journals and their subsets acting in conjunction with a contagious process.

The effect of the logarithmic transformation is similar for both *JACS* and *Science*. First, the location of the distributions as measured by the arithmetic mean shifts from near the bottom of the range to near the top of the range, indicating an increase in relative probability. Second, the variance is drastically below the arithmetic mean, resulting in extremely low variance-to-mean ratios—0.03 for *JACS* and 0.08 for *Science*. Third, instead of being skewed asymmetrically, the observations tend to distribute themselves symmetrically around the arithmetic mean within the constricted variance. This is the shape that results from random measurement error around the mean. From this demonstration it is easy to see that logarithmic transformation for purposes of inferential statistics results not in a more accurate description of reality, but is a mental model of reality artificially structured to conform to a law of error. It is interesting to note that the logarithmic transformation of the *JACS* distribution reveals *Science* as a possible outlier.

*4.2    Negative powerlaws at the level of the database*

While the previous analysis showed the lognormality of the distribution in a local citation environment, one can wonder whether this lognormality also exists in the larger dataset, that is, including the tails of the distributions. Is the *JCR* data loglinear? Does the logarithmic transformation provide us with a more adequate description of the citation distribution of these journals at the level of the database? Let us inspect the fit with a negative powerlaw by plotting the citation distributions of these 21 journals log-log using the full set of the 5907 journals included in this database.

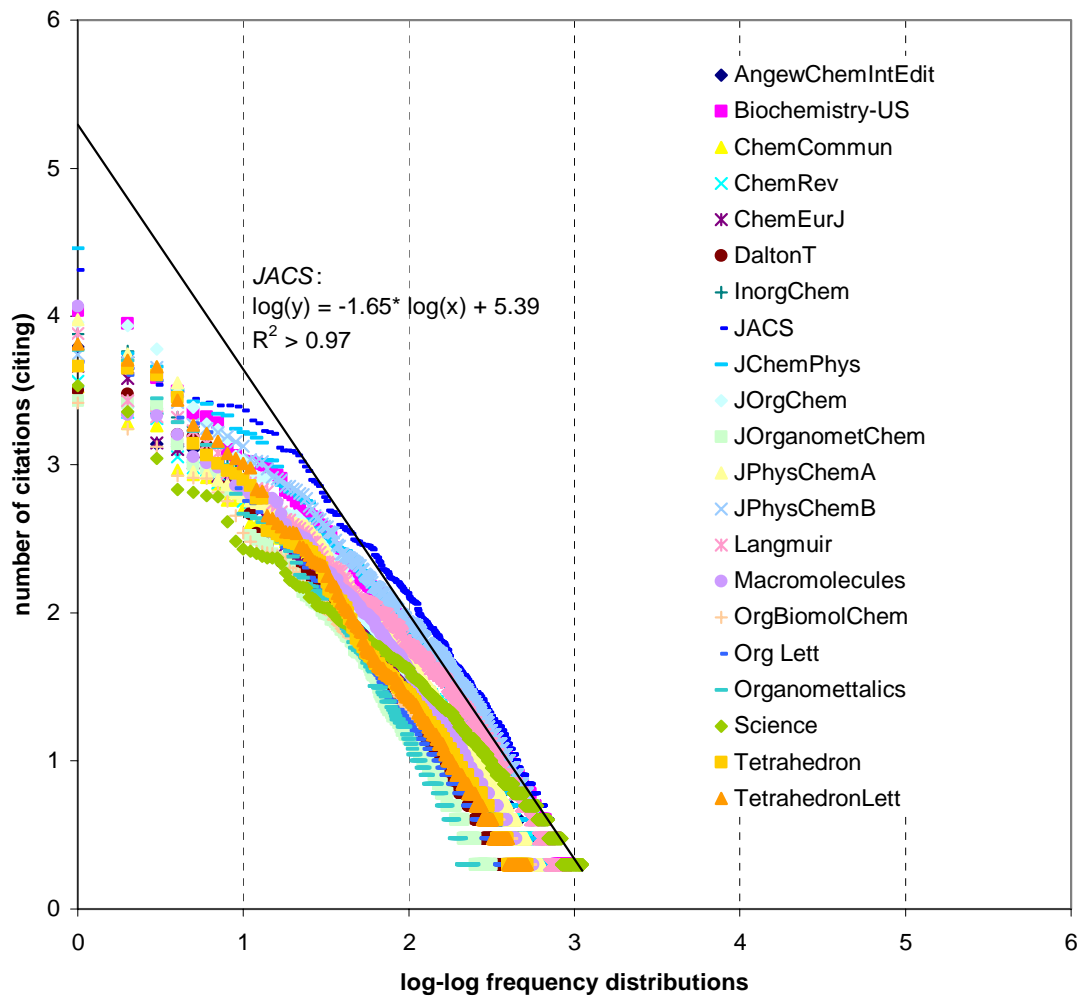

**Figure 3**: Citation distribution of 21 selected journals over the full journal set of 5907 journals included in the *JCR 2003*.

Figure 3 shows that the citation distributions of the journals exhibit the powerlaw-type distributions for the largest part of the curve (Barabási, 2002; Katz, 2000). The journals are related with citations to between $10^2$ and $10^3$ journals in their respective environments. (The number of journals in the *JCR* 2003 database was 5907.) The fits of the negative log-log curves are all high ($r^2 > 0.96$; see Table 2).

| Journal name | Number of journals in the citation environment | Citation distribution | Fit of log-log line |
|---|---|---|---|
| *Angew Chem Int Edit* | 686 | log(y) = -1.43 log(x) + 4.37 | $r^2 > 0.98$ |
| *Biochemistry-US* | 952 | log(y) = -1.53 log(x) + 4.89 | $r^2 > 0.97$ |
| *Chem Commun* | 500 | log(y) = -1.48 log(x) + 4.30 | $r^2 > 0.98$ |
| *Chem Rev* | 703 | log(y) = -1.49 log(x) + 4.51 | $r^2 > 0.97$ |
| *Chem-Eur J* | 530 | log(y) = -1.51 log(x) + 4.43 | $r^2 > 0.98$ |
| *Dalton T* | 394 | log(y) = -1.56 log(x) + 4.38 | $r^2 > 0.98$ |
| *Inorg Chem* | 558 | log(y) = -1.58 log(x) + 4.67 | $r^2 > 0.98$ |
| *J Am Chem Soc* | 981 | log(y) = -1.65 log(x) + 5.39 | $r^2 > 0.97$ |
| *J Chem Phys* | 728 | log(y) = -1.65 log(x) + 5.08 | $r^2 > 0.97$ |
| *J Org Chem* | 580 | log(y) = -1.64 log(x) + 4.80 | $r^2 > 0.98$ |
| *J Organomet Chem* | 315 | log(y) = -1.62 log(x) + 4.36 | $r^2 > 0.98$ |
| *J Phys Chem A* | 633 | log(y) = -1.56 log(x) + 4.71 | $r^2 > 0.97$ |
| *J Phys Chem B* | 869 | log(y) = -1.58 log(x) + 5.02 | $r^2 > 0.96$ |
| *Langmuir* | 892 | log(y) = -1.46 log(x) + 4.64 | $r^2 > 0.97$ |
| *Macromolecules* | 561 | log(y) = -1.58 log(x) + 4.65 | $r^2 > 0.97$ |
| *Org Biomol Chem* | 543 | log(y) = -1.39 log(x) + 4.08 | $r^2 > 0.99$ |
| *Org Lett* | 416 | log(y) = -1.58 log(x) + 4.39 | $r^2 > 0.97$ |
| *Organometallics* | 246 | log(y) = -1.78 log(x) + 4.65 | $r^2 > 0.98$ |
| *Science* | 1,113 | log(y) = -1.19 log(x) + 3.91 | $r^2 > 0.98$ |
| *Tetrahedron* | 518 | log(y) = -1.55 log(x) + 4.48 | $r^2 > 0.98$ |
| *Tetrahedron Lett* | 516 | log(y) = -1.59 log(x) + 4.55 | $r^2 > 0.99$ |

**Table 2**: Characterization of the powerlaw distributions the 21 selected journals

As has been noted before (Barabási *et al.*, 2002; Pennock *et al.*, 2002; Price & Thelwall, 2005), the initial parts of the distributions are typically 'hooked' off from the respective curves in the loglinear plots. Thus, there is a first environment of 20-50 journals which form a set with different relations with the journal under study than the larger set that fits the curve. This accords with the typical structure of specialties (20-50 journals) in which intellectually related journals cite each other more systematically than the larger set. The negative powerlaw fits to the scatter in the large tails of the distributions, but not to the core sets. The core sets follow a curvilinear distribution instead of a loglinear one.

In other words, nearby journals in the overall set experience *another* attraction to one another which is absent in their relations with more distanced journals. The latter pattern exhibits scattering, while the former pattern indicates the intellectual organization of these journals in specialties and fields. The deviation from loglinearity thus can be interpreted from the viewpoint of Brookes' model of Bradford's law as a very heterogeneous compound Poisson distribution and his conversion of the law into a linear model.  According to Brookes view of Bradford's law, this deviation from linearity is caused by the higher probability of the articles in the core set of journals to cite each other, while the remaining articles are spread over zones of journals that increase exponentially in number. The intellectual structure which organizes the data differently from the statistical expectation of loglinearity in the large tail of the distribution will be studied here below in order to see what the assumption of loglinearity would mean for retrieving structure in the intellectually organized core set.

*4.3     Factor analysis of the citation matrix*

Let us first input the citation matrix into a factor analysis without the logarithmic transformation. This analysis provides us with a baseline for assessing the effects of the logarithmic transformation in a next step. The so-called screeplot of the eigenvalues—which will be discussed below (Figure 5) in more detail because of the comparison with the transformed data—informs us that six-factors have an eigenvalue larger than unity. Table 3 provides this six factor solution. Factor designations were added in the second column using the LC scheme. (The factor loadings in a rotated component matrix are by definition equal to the correlation ($r$) of the hypothesized dimension with the variable.)

**Rotated Component Matrix(a)**

| ISI abbreviation for the journal name | Library of Congress classification | 1 | 2 | 3 | 4 | 5 | 6 |
|---|---|---|---|---|---|---|---|
| *Chem-Eur J* | Chemistry | .874 | .167 | -.126 | | -.122 | |
| *J Am Chem Soc* | Chemistry | .868 | .255 | .210 | | | .251 |
| *Chem Rev* | Chemistry | .866 | .356 | .195 | .173 | | |
| *Chem Commun* | Chemistry | .851 | .188 | -.230 | .117 | -.176 | -.237 |
| *Angew Chem Int Edit* | Chemistry | .749 | .150 | -.167 | | | |
| *Tetrahedron Lett* | Organic chemistry | .236 | .889 | -.127 | | -.169 | -.168 |
| *Tetrahedron* | Organic chemistry | .239 | .885 | -.117 | | -.176 | -.169 |
| *J Org Chem* | Organic chemistry | .323 | .876 | | | -.149 | -.130 |
| *Org Lett* | Organic chemistry | .386 | .838 | -.107 | | -.121 | |
| *Org Biomol Chem* | Organic chemistry | .158 | .387 | -.183 | -.333 | -.200 | |
| *Dalton T* | Inorganic chemistry | .423 | -.595 | -.199 | .313 | -.312 | -.337 |
| *Inorg Chem* | Inorganic chemistry | .564 | -.592 | | | -.283 | -.261 |
| *J Organomet Chem* | Organometallic chem. & compounds | .165 | -.103 | -.151 | .912 | -.168 | -.178 |
| *Organometallics* | Organometallic chem. & compounds | .271 | -.168 | -.104 | .901 | -.116 | |
| *J Phys Chem A* | Physical and theor. chemistry | | | .921 | | | |
| *J Chem Phys* | Physical and theor. chemistry | -.153 | -.118 | .872 | | | |
| *J Phys Chem B* | Physical and theor. chemistry | .118 | -.168 | .406 | -.157 | .602 | .212 |
| *Langmuir* | Surface chemistry | | -.144 | | -.138 | .808 | |
| *Macromolecules* | Polymers; macromolecules | -.180 | | | | .597 | -.140 |
| *Biochemistry-US* | Animal biochemistry | -.111 | | -.118 | | -.214 | .825 |
| *Science* | Science (general) | | -.199 | .144 | -.194 | .264 | .756 |

Extraction Method: Principal Component Analysis. Rotation Method: Varimax with Kaiser Normalization.
a Rotation converged in 9 iterations.

**Table 3**: Six factors explain 80.1% of the variance; factor designations added.

The fit with the classification of the Library of Congress is almost perfect. The only complication is the subclassification of the journals *Macromolecules*, *Langmuir*, and the *Journal of Physical Chemistry B.* This last journal is specifically indicated by our analysis as the journal which relates the specialties of physical and theoretical chemistry with surface chemistry and the study of polymers, while the hierarchical classification of the Library of Congress does not indicate this detailed pattern of relations.[8]

---

[8] Using the LC, *Macromolecules* can also be classified in QP801.P64 – Biochemistry, which has the following class hierarchy: Physiology—Animal biochemistry—Special substances—Organic substances—Miscellaneous organic substances, A-Z—Polymers. Macromolecules. This classification

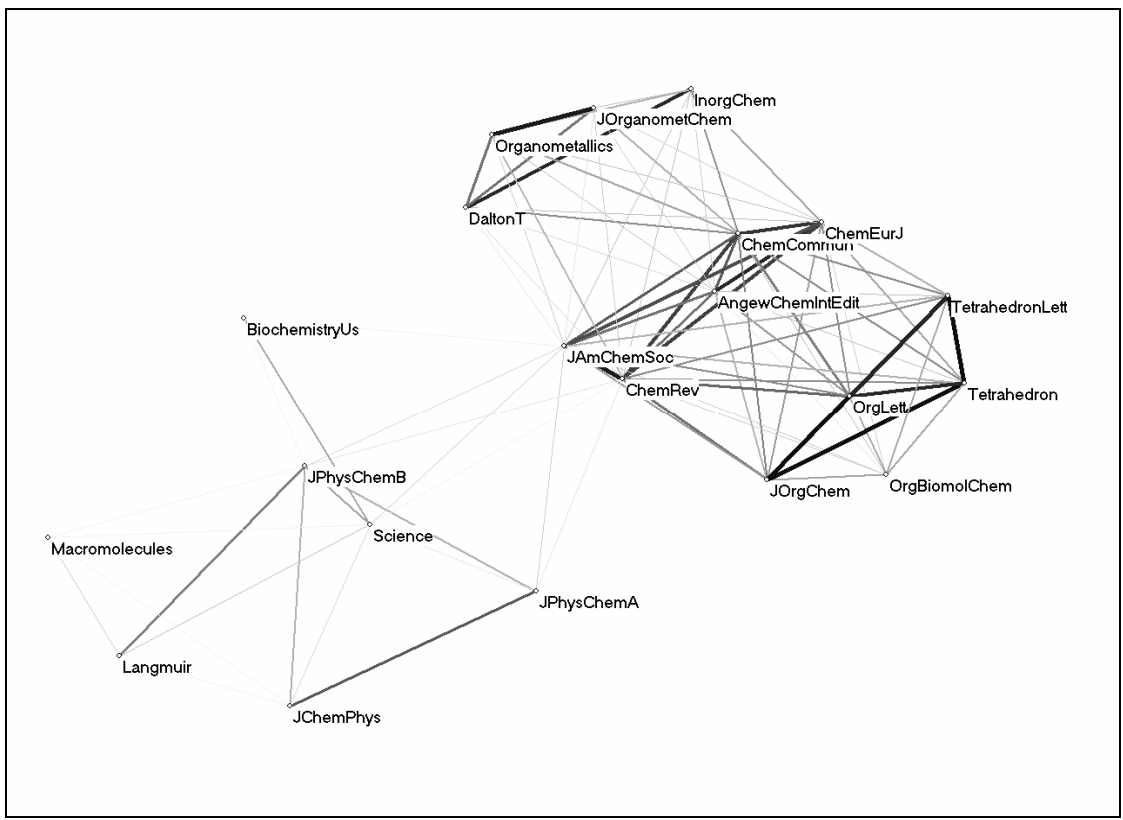

**Figure 4**: 21 journals in the citation environment of the *JACS* using Kamada & Kawai's (1989) algorithm on the basis of a Pearson correlation matrix ($r \geq 0$).

The visualization of the Pearson correlation matrix underlying the factor analysis (Figure 4) shows the groupings indicated by the factor analysis in considerable detail. For example, four core journals in organic chemistry form a strong bi-connected component ($r > 0.9$), while the journal *Organic and Biomolecular Chemistry* is related to this set at a lower level ($r > 0.5$). The major journals of chemistry are positioned in between organic and inorganic chemistry journals, and with variable relations to the physical chemistry group.

The journals *Science* and *Biochemistry-US* are classified as a separate group in this environment (factor 6), but with opposing signs of the loadings on factors 3 and 5

---

might make *Macromolecules* related to the journal *Biochemistry-US*, but this relation could not be retrieved using these citation-based methods.

which represent different subgroups of physical chemistry. While *Science* has a positive correlation with the physical chemistry set and to a lower extent with the set of general chemistry journals,[9] the citation pattern of *Biochemistry-US* has a negative correlation with all these sets. Neither the citation pattern of *Biochemistry-US* nor that of *Science* shows significant correlation with any of the other journals in the set. These two journals are drawn into the citation environment of the *JACS* as members of a relational graph among large journals. The two journals are grouped together because they share this relation with the *JACS* in a next-order network.

*4.4    Log-scaled matrix*

In a second step we proceed by applying the transformation of taking the logarithm of all cells in the matrix. This reduces the variance in the matrix enormously. Four factors instead of six now have an eigenvalue larger than unity. Figure 5 shows the scree plots for the distributions of the eigenvalues before and after the transformation. In other words, the transformation reduces the dimensionality in terms of eigenvectors in the matrix in addition to the variance in the data. This effect may be counter-productive if one wishes to distinguish statistically among the groupings.

---

[9] For reasons of presentation factor loading < 0.1 are not exhibited in Table 3.

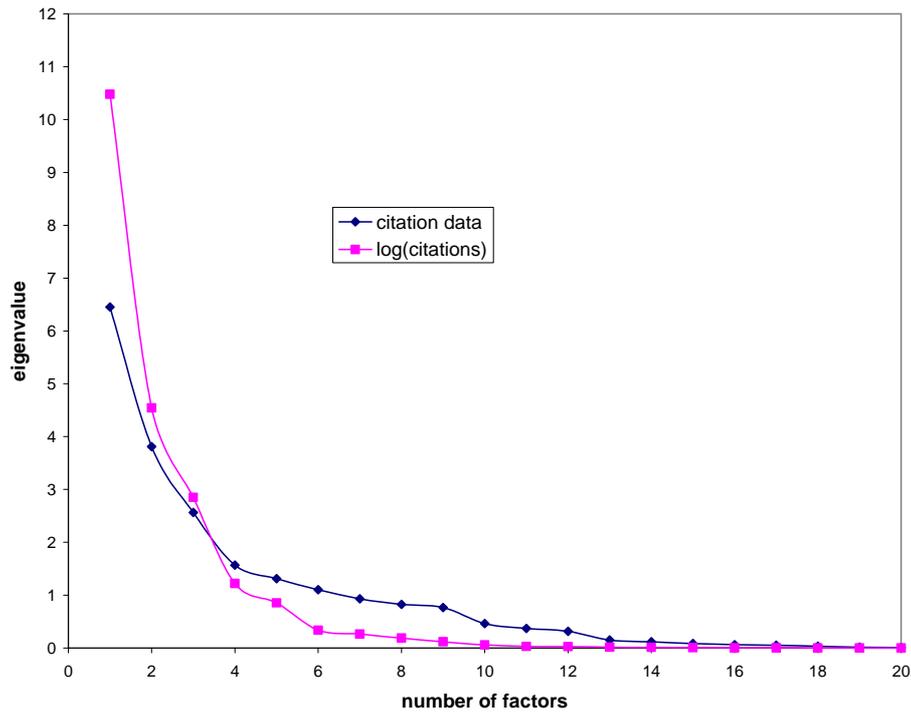

**Figure 5**: Screeplots of citation patterns before and after the logarithmic transformation.

The four-dimensional rotated factor solution (explaining 90.0% of the variance and the default in SPSS) classifies the journals *Science* and *Biochemistry-US* as belonging to the cluster of 'physical chemistry' journals. Although this may already count as an argument against the logarithmic transformation, let us give the opposing argument the benefit of the doubt by deliberately forcing six factors as in the untransformed case. A larger number of factors enhances a further differentiation of the grouping (Table 4).

**Rotated Component Matrix(a)**

| ISI abbreviation of the journal name | Library of Congress classification | 1 | 2 | 3 | 4 | 5 | 6 |
|---|---|---|---|---|---|---|---|
| *Tetrahedron* | Organic chemistry | .953 | .165 | -.166 | -.163 | | |
| *J Org Chem* | Organic chemistry | .951 | .214 | -.172 | -.104 | | |
| *Org Lett* | Organic chemistry | .948 | .175 | -.104 | -.117 | -.122 | |
| *Tetrahedron Lett* | Organic chemistry | .937 | .212 | -.154 | -.138 | -.142 | |
| *Org Biomol Chem* | Organic chemistry | .843 | | .262 | -.175 | -.215 | |
| *Dalton T* | Inorganic chemistry | | .980 | -.139 | | | |
| *Inorg Chem* | Inorganic chemistry | | .973 | | | .180 | .105 |
| *Organometallics* | Organometallic chem. & compounds | .366 | .787 | -.425 | | | |
| *J Organomet Chem* | Organometallic chem. & compounds | .459 | .719 | -.508 | | | |
| *J Am Chem Soc* | Chemistry | .651 | .527 | .155 | .187 | .337 | .312 |
| *Chem Rev* | Chemistry | .631 | .558 | | .270 | .238 | .356 |
| *Chem-Eur J* | Chemistry | .675 | .685 | -.132 | .129 | | |
| *Chem Commun* | Chemistry | .653 | .683 | -.165 | .135 | | |
| *Angew Chem Int Edit* | Chemistry | .669 | .673 | -.148 | .137 | | -.171 |
| *Biochemistry-US* | Animal biochemistry | | -.181 | .935 | | .183 | .104 |
| *Science* | Science (general) | -.345 | -.281 | .699 | .305 | .230 | -.267 |
| *J Phys Chem A* | Physical and theoretical chemistry | | .130 | .203 | .130 | .957 | |
| *J Chem Phys* | Physical and theoretical chemistry | -.335 | | .147 | .415 | .821 | |
| *J Phys Chem B* | Physical and theoretical chemistry | -.189 | | .589 | .575 | .503 | |
| *Macromolecules* | Polymers; macromolecules | -.133 | .104 | -.120 | .934 | .174 | .131 |
| *Langmuir* | Surface chemistry | | -.106 | .416 | .843 | .231 | -.162 |

Extraction Method: Principal Component Analysis. Rotation Method: Varimax with Kaiser Normalization.
a Rotation converged in 6 iterations.

**Table 4:** Six factors explain 96.6% of the variance; factor designations added.

The two factor solutions (before and after the transformation) do not lead to an essentially different classification, but the order of the factors is different and some groupings are less pronounced after the transformation. For example, the two journals belonging to 'organometallic chemistry' are after the transformation subsumed under the group of two 'inorganic chemistry' journals, albeit with different loadings on other factors. In the previous case these two journals of organometallic chemistry spearheaded factor 4 as a separate dimension. Furthermore, the journals *Biochemistry-US* and *Science* are not demarcated from the group of physical chemistry journals with which they now share factor loadings on several dimensions. In the case of the journal *Biochemistry-US* this is completely mistaken according to the results of both the analysis of the untransformed data and the LC classification.

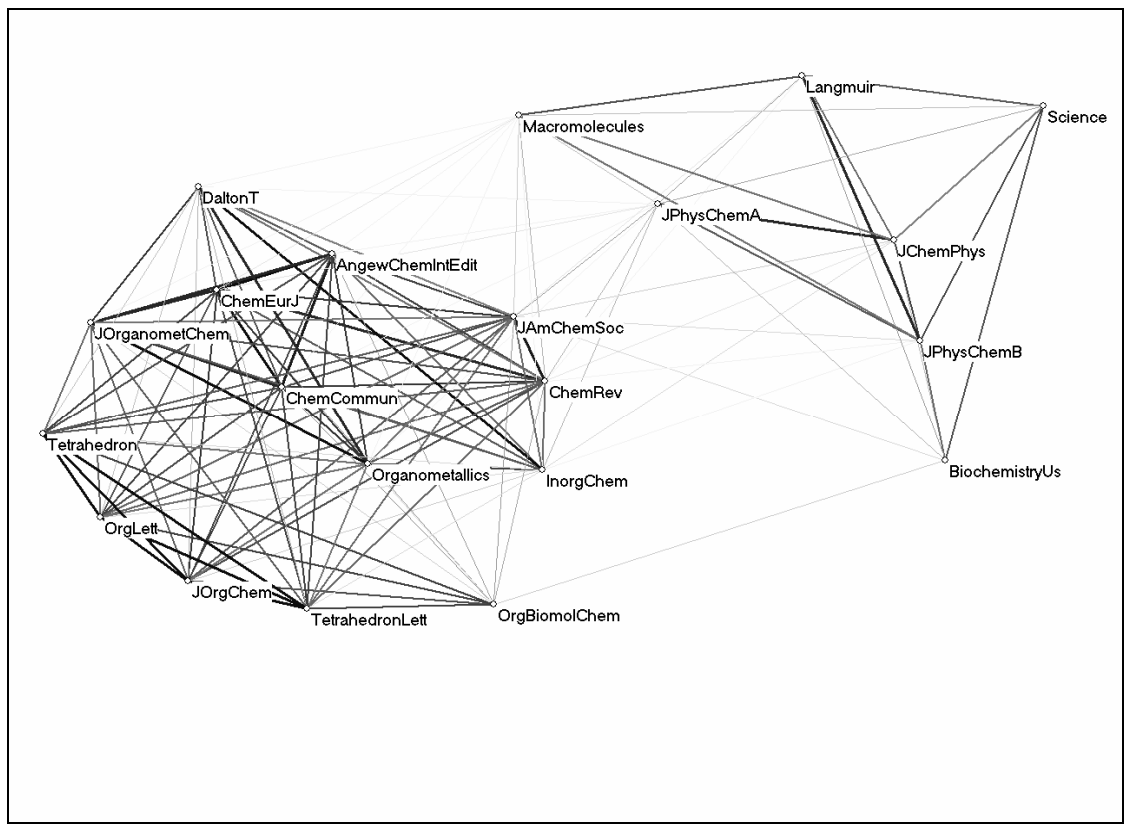

**Figure 6**: Relations among the logarithms of the citations of 21 journals in the citation environment of *JACS* (*JCR* 2003; $r \geq 0$).

The visualization of the (positive) Pearson correlations (Figure 6) no longer explains the structure in the data. The two journals in organometallics are now providing the interface between the organic chemistry journals, on the one side, and the common grouping of the inorganic chemistry and general chemistry journals on the other. The previous factor solution taught us that the general chemistry journals share more communality with the organic chemistry set than with the inorganic chemistry journals. However, this picture suggests that the general chemistry journals provide a focus within the inorganic chemistry set.

*Science* was included as part of the set because it played a sufficiently important role in the wider citation environment of the *JACS*. Like *Biochemistry-US*, *Science* plays a role in a network of large journals surrounding the *JACS*. However, the positive correlation between the citation patterns of *JACS* and *Science* ($r = 0.232$) turns negative ($r = -0.187$) after the logarithmic transformation. Consequently, *Biochemistry-US* has become even more closely related to the physical chemistry group of journals than *Science* after the transformation. The intellectual differentiation among these dimensions of the data set is thus distorted by the logarithmic transformation.

Nevertheless, one could argue that the structure in the data remains in many respects robust against the distortion produced by the logarithmic transformation. Important elements of the previously retrieved structure in the data could still be found after the transformation. While the rotated factor analysis is robust against the log-based transformation, the vector-space model used for the visualization was not.[10] The structure in the database is suppressed and it becomes more difficult to distinguish the relevant delineations.

**5. Discussion**

The logarithmic transformation was primarily needed in order to restore the assumption of normality in the distributions underlying the Pearson correlation.

---

[10] The vector-space model is usually associated with using the cosine (Salton & McGill, 1983), but because of the equivalence between the cosine and the Pearson (Jones & Furnas, 1987) the concept of a vector-space can be associated equally well with the Pearson correlation matrix.

However, factor analysis itself does not require these distributional assumptions. For example, Kim & Mueller (1978, 74f.) note that even ordinal data can be used for the factor analysis. Pearson correlations, however, will be attenuated when variables come from a variety of underlying distributions. As we have seen, the factor analysis may be robust nevertheless.

When one only needs to consider similarity (e.g., for the visualization) and no further statistics are required, the sensitivity of the Pearson correlation to zeros and outliers may be considered as a reason for using the cosine as a measure instead (Ahlgren *et al.*, 2003). While the cosine is not a statistical measure, it allows for a more precise appreciation of the outliers (and zeros) in the distribution, exceptions for which the logarithmic transformation precisely tried to correct. In our opinion, both the cosine and the Pearson correlation are valid similarity measures; the difference is only in the *a priori* normalization to the mean (Jones & Furnas, 1987). This can be an advantage or a disadvantage depending on the research question.

Let us consider analytically how these two similarity measures are affected by the logarithmic transformation by providing a stylistic example. Assume a logarithmic series like 1, 10, 100, 1000 in one variable (v1), and another variable (v2) in which the two top values are reversed as follows: 1, 10, 1000, 100. The corresponding variables log(v1) and log(v2) would thus read 1, 2, 3, 4 and 1, 2, 4, 3, respectively.

|  | *v1 versus v2* | *log(v1) vs log(v2)* |
|---|---|---|
| *Pearson's r* | −0.155 | +0.800 |
| *Cosine* | +0.198 | +0.967 |

**Table 5**: Effects of the logarithmic transformation on two variables v1 and v2 using the three similarity criteria.

The Pearson correlation between the original variables v1 and v2 is negative (r = −0.155), while the correlation between the logarithmically transformed varibles log(v1) and log(v2) is +0.800. Thus, the sign of the correlation is changed. However, the effect on the cosine would be even more dramatic: the cosine between log(v1) and log(v2) is +0.967 as against a relatively low value for the cosine of +0.198 for the comparison between v1 and v2. This means that the two distributions are considered as virtually similar after the logarithmic transformation; they are no longer distinguishable in terms of the vector-space model because the value of the cosine is very close to unity.[11]

The logarithmic transformation obscures the outliers and therefore the differences among the distributions. We have seen a similar change in the sign of a correlation above for the empirical case of the network among major journals like *Science, Biochemistry-US,* and the *JACS*. Thus, the reduction of the variance by the logarithmic transformation corrupts the structural elements in the metrics of the network which are interesting for the classification. The transformation not only reduces the variance, but also the latent structure underlying the variance. Structural differences should not be reduced on *a priori* grounds if one wishes to reveal structural dimensions by means of analytical techniques.

---

[11] Since the cosine is not based on a normalization, the Cartesian space is spanned from the perspective of the origin. Any reduction of the variance will lead to higher values for the cosines from this perspective external to the system. Thus, the effects of the logarithmic transformation on the Pearson correlations are further enhanced for the cosine as a similarity measure. However, the logarithmic transformation is not pertinent to the cosine because this measure provides no basis for probabilistic inferences.

**6. Conclusions**

A network of communications can be analyzed in terms of its eigenvectors, that is, dimensions or factors. Although hierarchy can be expected to prevail in each of the dimensions, the dimensions can become increasingly differentiated in their relations to one another because the variety of the dimensions enables the system to process more complexity. In the factor-analytic model the dimensions are usually assumed to be orthogonal. Since the model is an idealization, covariations among the dimensions can also be expected. One can also formulate this in terms of systems theory as the expectation of *near* decomposability in the organization of complex systems (Simon, 1973).

For example, the sciences—disciplines, specialties, etc.—operate mainly in parallel to one another. Citation densities are high within units and much lower among them. In order to identify the eigenvectors in the networks of communication, the outliers provide us with a focus and the off-diagonal zeros support the decomposability of the matrix. Thus, these extreme elements have the crucial function of spanning the multi-dimensional space. The sensitivity of the model for outliers and zeros is a desired feature in this case. If one is interested in revealing the different dimensions of the structure, the *a priori* reduction of the variance by a logarithmic transformation can be counter-productive. From the perspective of the descriptive statistics, one is interested precisely in the curvilinear parts of the curves where the distributions deviate from the loglinear or powerlaw-like distributions because one may be able to hypothesize substantive reasons for the deviations (Ferrer Cancho & Solé, 2001; Pennock *et al*., 2002). Note that these deviations were statistically insignificant from the perspective of the fit to the negative powerlaw (Table 3) because the citation patterns of the larger number of journals in the tail can be expected to fit almost perfectly with this curve (Katz, 1999, 2000).

From the perspective of inferential statistics, the outliers can be considered as errors, but for the analysis of structure these deviations from the powerlaw-type distributions are essential information. The intellectual organization of the scientific journals into next-order structures like specialties and disciplines generates the heterogeneity and the compoundedness of the distributions because each of these structural elements can be expected to have specific publication and citation characteristics. Once the structural dimensions have been determined, for example, by using the technique of factor analysis, these dimensions constitute a second-order variation which can be taken as input for inferential statistics. For example, one can use the factors as latent variables in a structural equation model (Jöreskög & Goldberger, 1975; Bray & Maxwell, 1985, pp. 61 ff.; Leydesdorff, 1995, p. 57f.).

The logarithmic transformation did not contribute to clarification in the case of our relatively robust set of aggregated journal-journal citation data, but it did also not completely ruin the underlying factor structure. Aggregated journal-journal citation relations provide relatively robust structures which are reproduced from year to year to a considerable extent (Leydesdorff, 2002). Had we used word-pattern distributions in texts (e.g., titles or keywords) as data, this assumption of reproducibility over time would no longer hold true (Leydesdorff, 1997). However, at each moment in time, the outliers are structuring the systems under study. The factor analysis (based on rotating a Pearson correlation matrix) can thus remain useful for the classification at each moment. However, one would expect an even more drastic reduction of explanatory power for the prediction of underlying structure if logarithmic transformation is applied in the case of less robust datasets.